\documentclass[11pt,twoside]{article}
\usepackage{./asp2010,graphicx}

\resetcounters

\begin{document}

\title{Using Gas Kinematics To Constrain 3D Models of Disks: IC 2531}
\author{Arthur~Eigenbrot$^1$, Matthew~Bershady$^1$
\affil{$^1$University of Wisconsin Madison, 475 N. Charter Street, Madison, WI 53706}}

\begin{abstract}
We use deep, longslit spectra of the nearby edge on galaxy IC 2531 to obtain
gas kinematics out to 5 radial scale-lengths (40 kpc) and 4
vertical scale-heights (1.7 kpc). The large
vertical range spanned by our data offers unique leverage to constrain
three-dimensional models. The shape of the observed emission-line profiles offer insights to line-of-sight density distributions in the disk, and we discuss the possibility that we are
seeing disk-flaring in the ionized gas. Finally, we begin to quantify measurements of line shape to allow model galaxies to be compared to data across all radii and heights simultaneously.
\end{abstract}

\section{Introduction}
Spectroscopic studies of edge-on galaxies are a useful tool for understanding the vertical stratification of gas and stellar kinematics as well as population gradients in disks. We are undertaking a program to characterize the vertical structure of galaxy disks to verify and refine recent estimates of disk sub-maximality in face-one surveys \citep{bershady11,DMSIII,Martinsson13b}. As part of this program we have acquired optical kinematic data of a single edge-on galaxy, IC 2531 (ESO 435-G25), over a large range in radius and at multiple heights above the mid-plane. In addition to addressing the question of disk-maximality these data can provide insights to line-of-sight density distributions not normally accessible to purely face-on studies.

Previous attempts to use kinematic measurements to measure line-of-sight structure in edge-on galaxies have relied on HI observations \citep{KregII,Oosterloo07}, where the high spectral resolution of radio data allows detailed line profile shapes to be measured. Unfortunately, the \emph{spatial} resolution of such studies is too low to allow for meaningful comparison to models of the disk mid-plane structure. Similar studies in the optical have been made \citep{KregIV}, but only at a single height. Our measurements of IC 2531 provide a unique opportunity because 1). they are taken at high enough spectral \emph{and} spatial resolution to allow for the study of detailed line shapes across a wide, and finely sampled, range of radii; and 2). they span a range of galactic heights, which will allow us to estimate a fully 3D density distribution for this galaxy. In particular, recent studies of face-on galaxies \citep{DMSIV,Martinsson13a} indicate that the vertical velocity distribution of gas, $\sigma_{z, gas}$, is constant with radius, while the same quantity for stars, $\sigma_{z,\ast}$, decreases exponentially with radius. This implies that the emitting gas occupies a flared disk, and our off-plane data is well suited to testing this hypothesis.

\section{Data and Line Profile Models}
The data for IC 2531 were obtained using the Robert Stobie Spectrograph \citep[RSS;][]{RSS} on the Southern African Large Telescope (SALT) during 2011 Semester 3. Kinematic emission line data were obtained from the center of the galaxy out to $R=\pm 5h_r$ and $z= 0, 1, 2,$ and $4 h_z$, where $h_r$ and $h_z$ are the radial and vertical exponential scale lengths of the stellar disk.

To model the shape of emission line profiles we first construct a full, 3D galaxy model using custom software. The models contain a geometric prescription for the distribution of emitting gas, an independent description for the distribution of absorbing dust, a dynamical model, and an accounting of radiative transfer along any line of sight.  The specific arrangement of gas and dust depends on what morphologies are present in the model (flares, spiral arms, etc.), but all models begin with a smooth, doubly-exponential disk; different morphologies are then achieved by redistributing the light from the basic disk model. In this way the total amount of gas and dust does not change, regardless of what morphologies are present, which simplifies comparisons between models. The model rotation curves are parametrized as $V_c(R) = V_{c,0}\mathrm{~tanh}(R/h_{r,dyn})$, where $h_{r,dyn}$ is the dynamical scale length \citep{Andersen13}. The parametrization for spiral arms is based off that of \citet{ASRSpiral}.

Once a model galaxy has been constructed we ``observe'' it along multiple edge-on sight lines. Each point along a sight line is assigned a weight based on its surface brightness and optical depth. These weights are then used, in conjunction with the tangential velocity of each point, to construct an apparent line-of-sight velocity distribution function (LOSVDF). We then place emitters at each point and given them a characteristic line width of 17 km/s. The final line profile is computed by summing the contributions of all the emitters (weighted by the apparent LOSVDF) and simulating the effects of the instrument setup used to take data of IC 2531 (SALT/RSS).

We constructed models for three different galaxy morphologies: a smooth, doubly-exponential disk, a smooth disk with a flare, and a disk with spiral arms.

\section{Results}
\articlefiguretwo{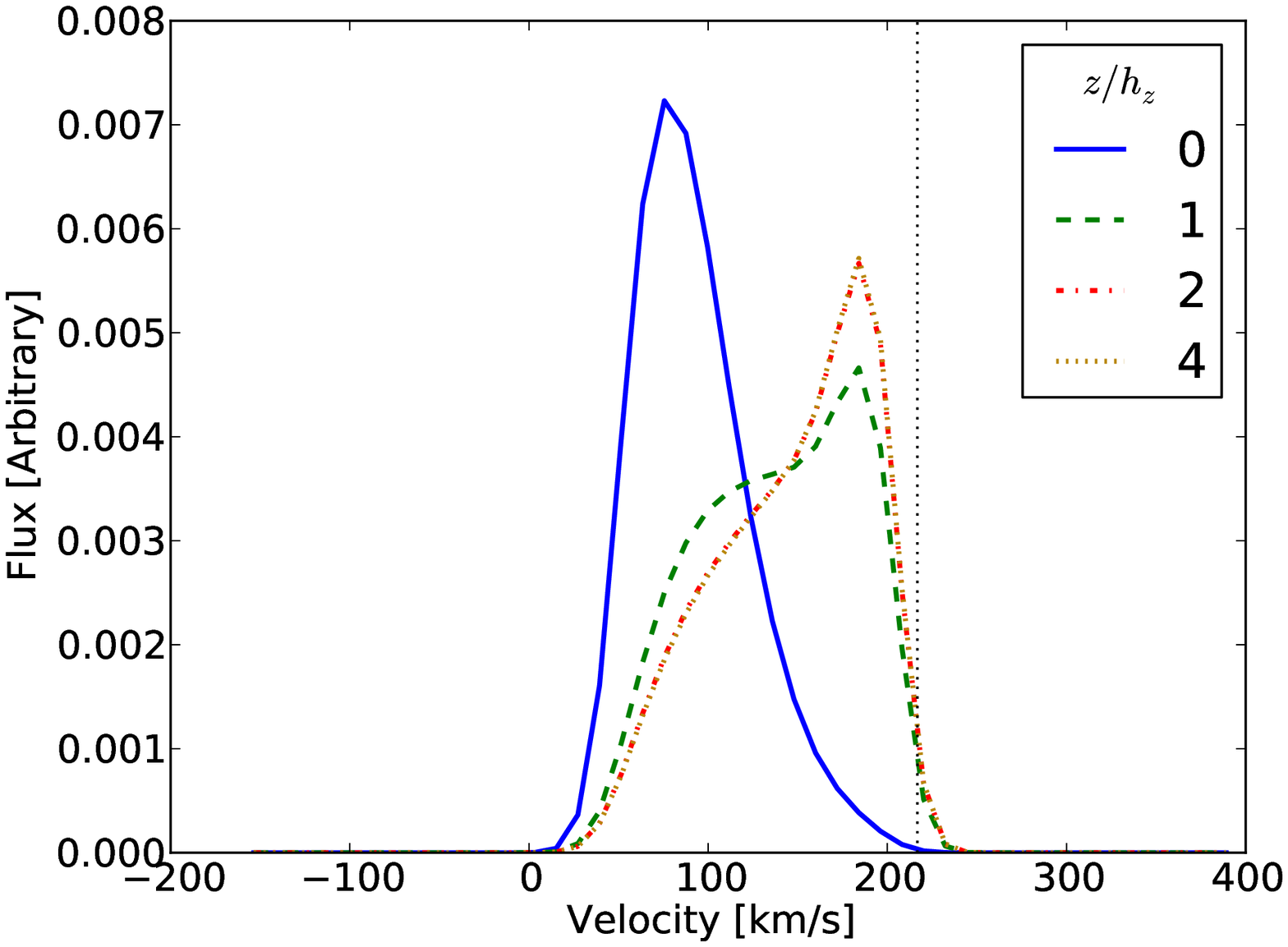}{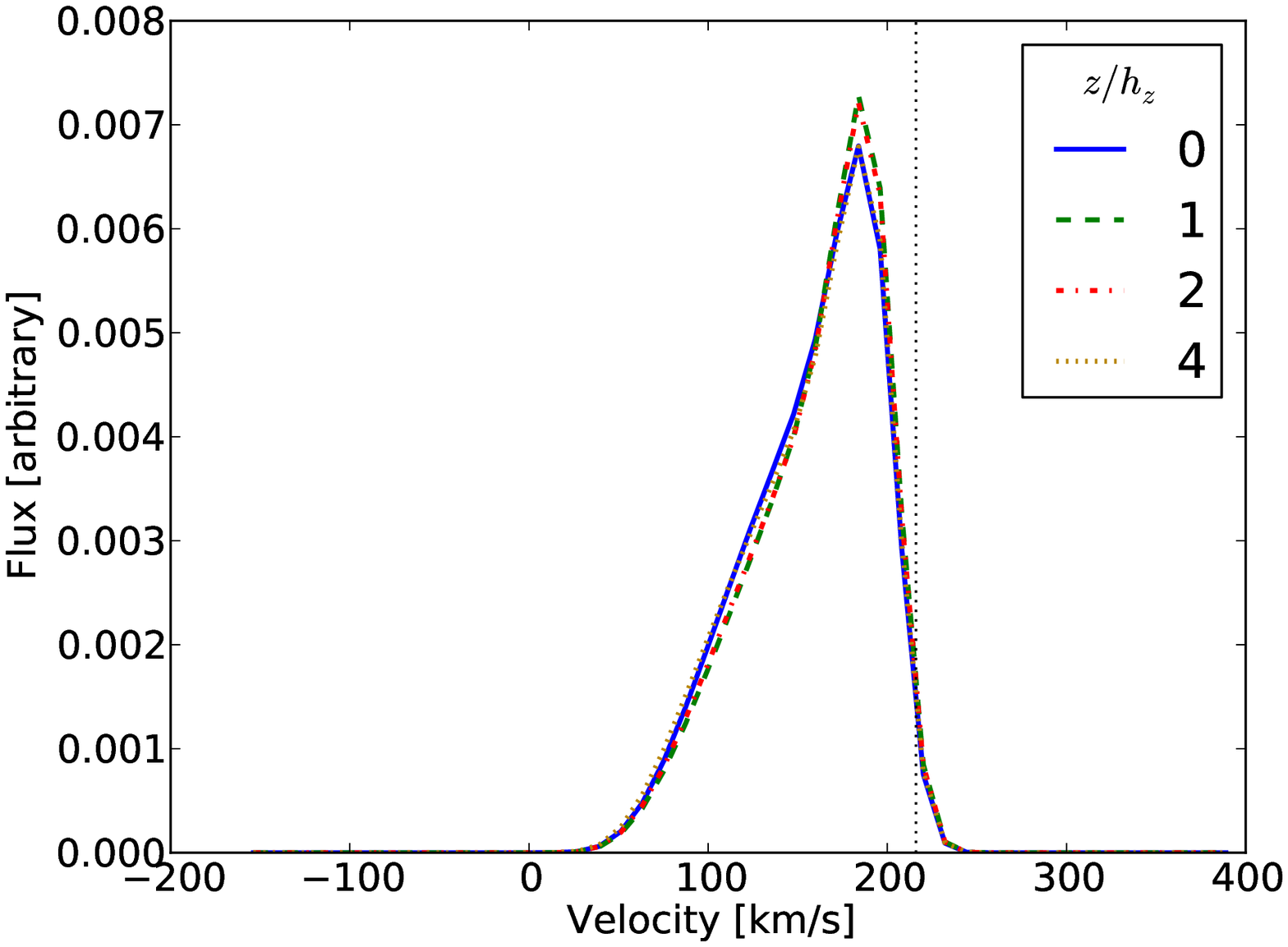}{fig:profiles}{Model line profiles at four heights above the mid-plane for a smooth, doubly-exponential disk (left) and a smooth disk with a flare (right). The vertical dotted lines show the true circular speed, $V_c$ (217 km s$^{-1}$). Notice that the sign of the skew of the line profiles changes with height in the non-flared model, but does not change in the flared model.}

Figure \ref{fig:profiles} shows an example of how different disk morphologies affect the shape of emission line profiles. In particular, in a smooth, doubly-exponential disk model the skew of a line profile changes sign (from positive to negative) as height is increased while a flared disk model produces profiles with negative skew at all heights. This is due to a redistribution of dust by the flare that allows sight-lines to penetrate farther into the galaxy at low heights.

\articlefiguretwo{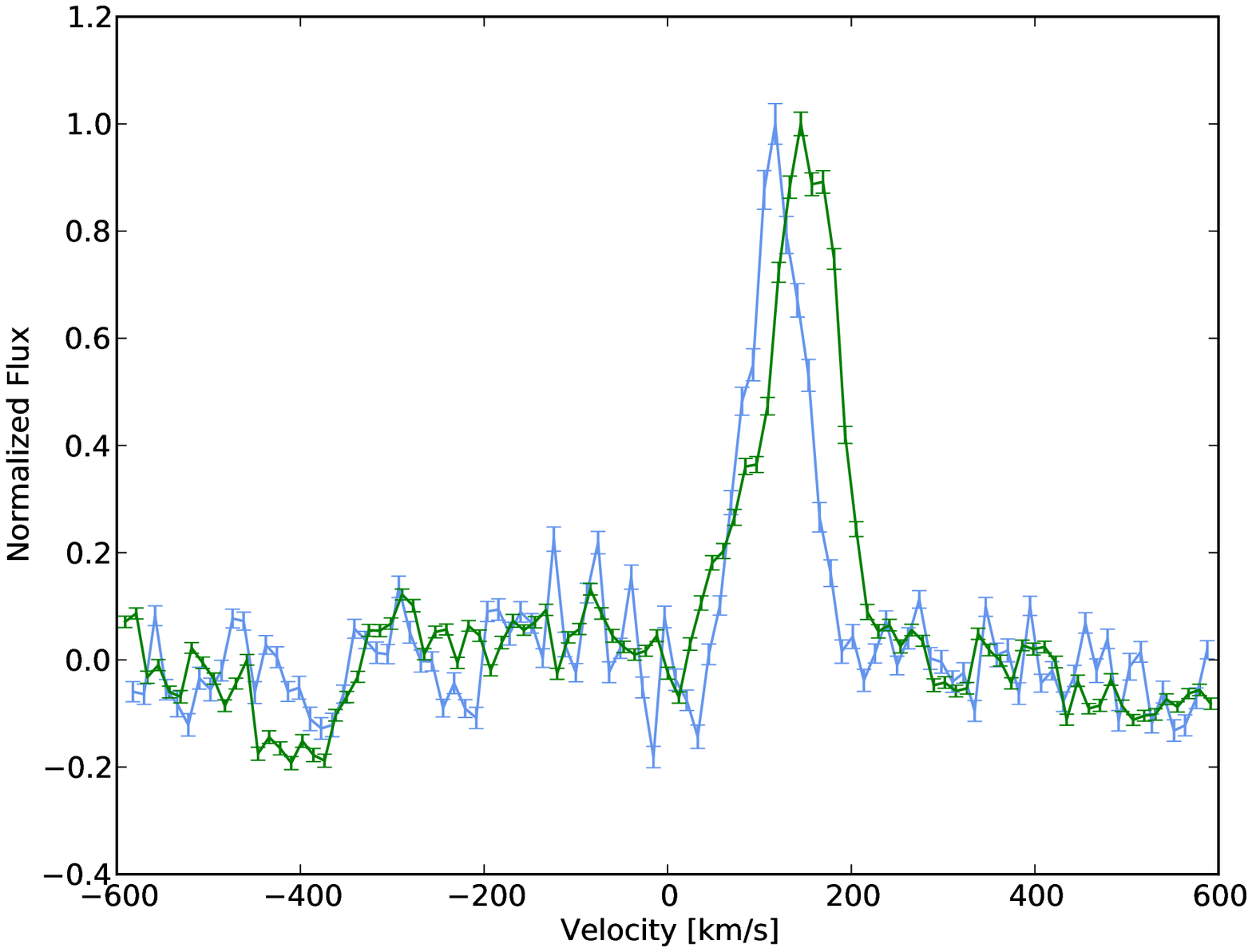}{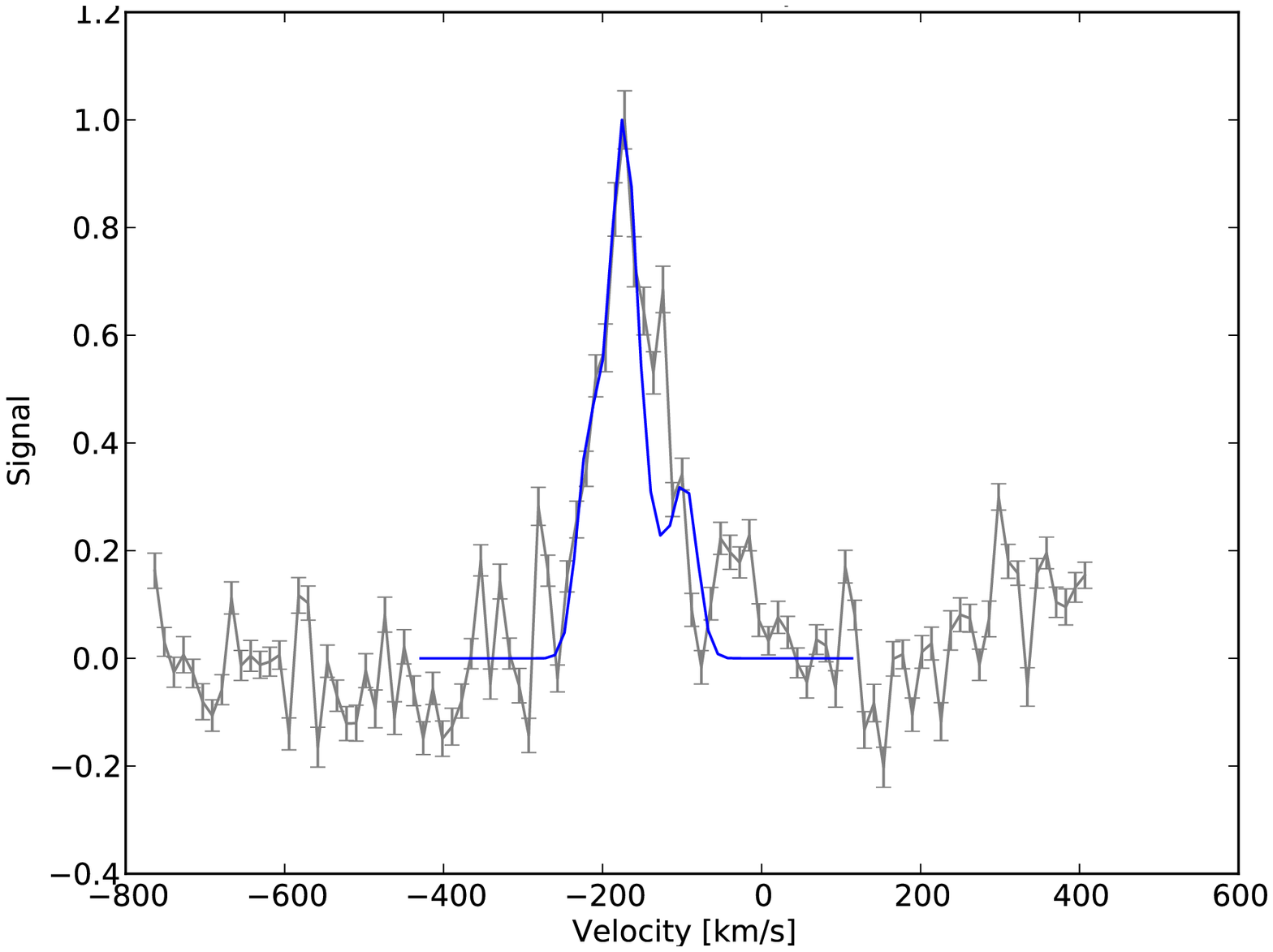}{fig:data}{\emph{Left:} [OIII]$_{5007}$ emission lines from IC 2531 at the same radius $(\sim 1 h_r)$ but at two different heights ($0h_z$, blue, and $1h_z$, green). Both of these profiles have a negative skew. \emph{Right:} \emph{Grey:}Measured [OIII]$_{5007}$ emission line at $(R,z)=(\sim 2 h_r,0)$. \emph{Blue:} Model line profile with spiral arms. The model was constructed so that the location of the peak of this emission line matches the data.}

Figure \ref{fig:data} shows example [OIII]$_{5007}$ emission lines from IC 2531. The left plot of Fig. \ref{fig:data} shows data at the same radius $(\sim 1 h_r)$ but different heights $(0h_z, 1h_z)$. Notice that both lines have a negative skew. This is in direct disagreement with the predictions of a smooth, doubly-exponential disk model (Fig. \ref{fig:profiles}) and supports the theory that the emitting gas occupies a flared disk. The right plot of Fig. \ref{fig:data} shows how well a spiral arm model matches an observed emission line. We find similarly good fits between a spiral arm model and our data at multiple radii. The impact of spiral arms is immediately visible in the model line-profile; each sub-peak  corresponds to the sight line crossing a spiral arm. Using this information it is possible to deconstruct the line-of-sight density distribution simply by looking at the spatial coherency of detailed line shapes. 

\articlefiguretwo{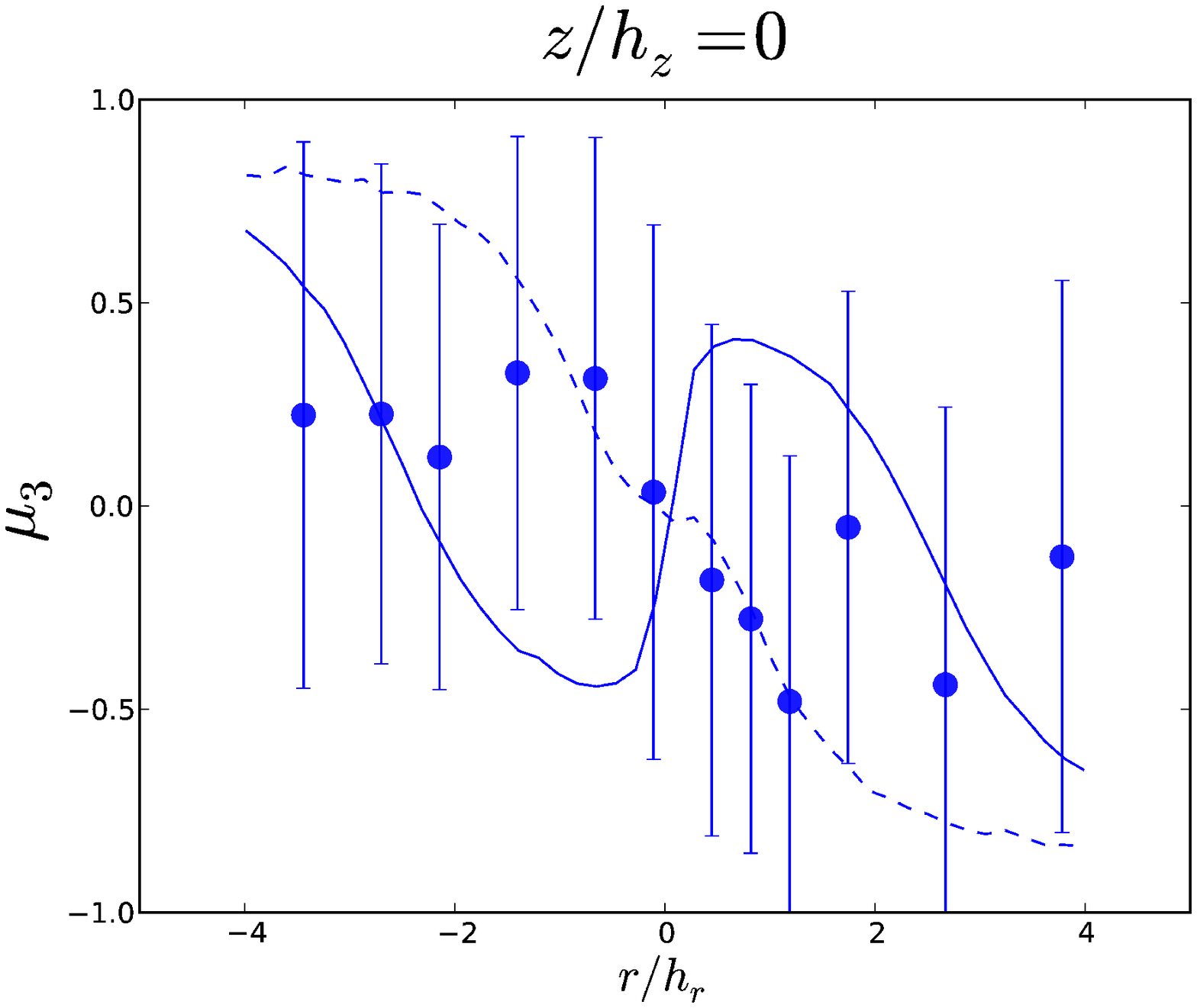}{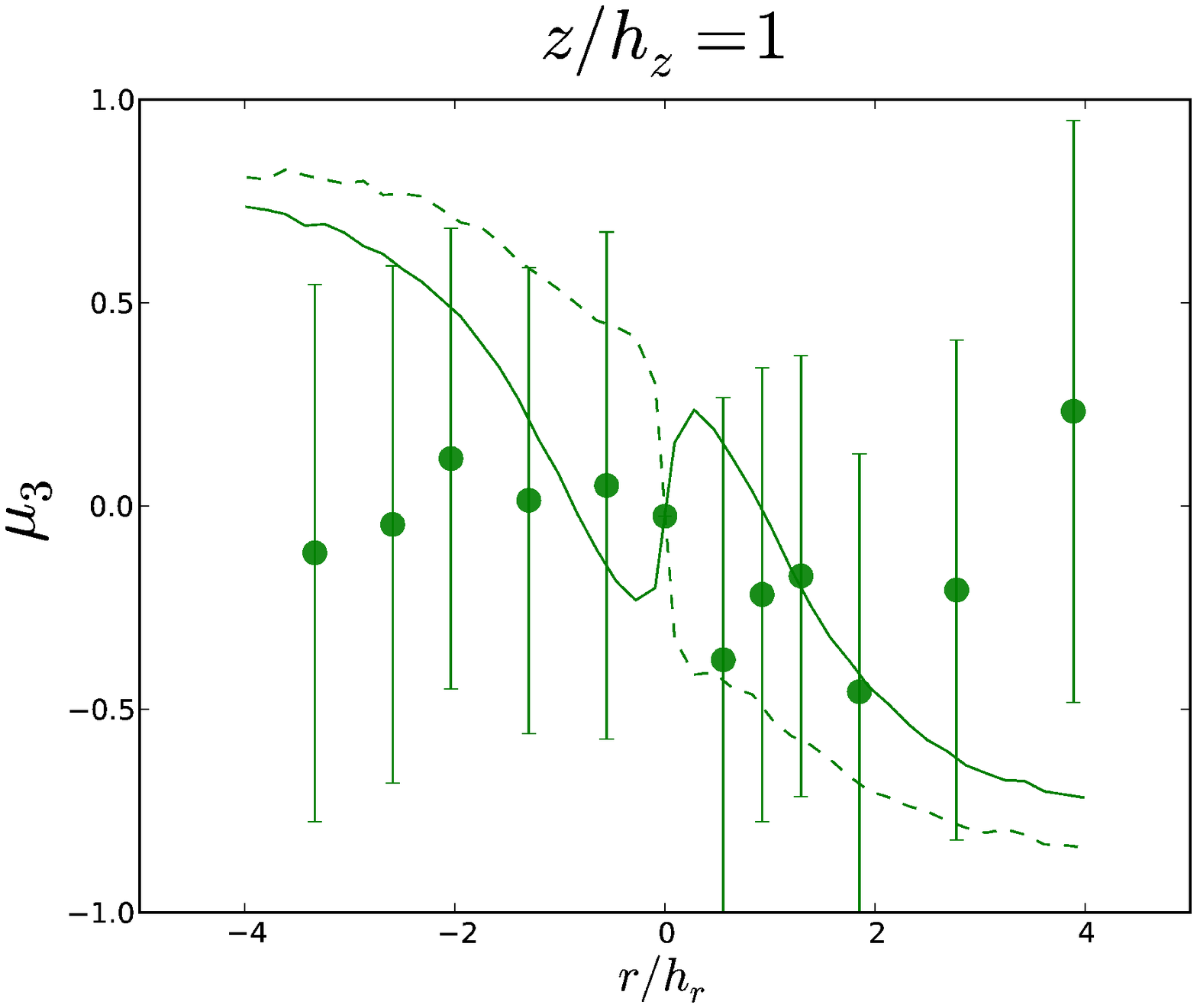}{fig:moments}{Skew ($\mu_3$) as a function of radius at four different heights. The points are measurements from our [OIII]$_{5007}$ data of IC 2531. The lines show the same measurements for galaxies models with a smooth, doubly-exponential disk (\emph{solid}) and a flared disk (\emph{dashed}).}

We have started an effort to quantify how well model line shapes match our data over a large range of radii and heights, focusing first on constraining the large-scale disk structure, such as flares (as opposed to spiral arms). Simple $\chi^2$ tests can tell us how well a single pair of model and data lines match, but they are highly sensitive to slight mismatches in the peak location and do not allow for robust comparison between models and data that constrains large-scale disk structure.

To address these issues we use statistical moments about the mean as quantifiers of line shape; we can measure the shape of both model and data lines and compare how the shape changes with radius and height. Figure \ref{fig:moments} shows an example of this using skew ($\mu_3$) as the line-shape indicator. Neither smooth, doubly-exponential or flared disk models fit the data at all radii and heights, although the flared-disk model does a better job describing line-skew of the inner disk near the mid-plane. Current measurement errors are over-estimated, but future refinements and the simultaneous use of multiple moments will enable us to better constrain a wider range of models. Extension of this technique to fitting the detailed line-profile shape over all heights and radii will be used to map spiral structure.

\acknowledgements This work is supported by NSF grant AST-1009471. All of the observations reported in this paper were obtained with the Southern African Large Telescope (SALT), proposal 2011-3-UW-002 (Bershady). We would also like to acknowledge L. Matthews for discussion which stimulated this work.

\bibliography{eigenbrota}

\end{document}